\documentstyle[12pt]{article}
\begin{document}

\begin{center}
\centering{{\bf ABSOLUTELY STABLE SOLITONS IN TWO-COMPONENT ACTIVE SYSTEMS}}
\end{center}

\vspace*{1.0cm}
\begin{center}
Boris A. Malomed\footnote{electronic address malomed@math.tau.ac.il;
the author to whom correspondence should be sent}
\end{center}

\begin{center}
\centering{
Department of Applied Mathematics,
School of Mathematical Sciences,
Tel Aviv University,
Ramat Aviv 69978, Israel}
\end{center}
\begin{center}
Herbert G. Winful
\end{center}
\begin{center}
Department of Electrical Engineering and Computer Science,
University of Michigan, Ann Arbor, Michigan 48109-2122
\end{center}

\newpage
\begin{center}{\large\bf ABSTRACT}\end{center}
As is known, a solitary pulse in the complex cubic
Ginzburg-Landau (GL) equation is unstable.
We demonstrate that a system of two linearly coupled GL
equations with gain and dissipation in
one subsystem and pure dissipation in another produces
absolutely stable solitons and their bound states. The problem is solved in
a fully analytical form by means of the perturbation theory. The soliton
coexists with a stable trivial state; there is also an unstable soliton
with a smaller amplitude, which is a separatrix between the two stable
states. This model has a direct application in nonlinear
fiber optics, describing an Erbium-doped laser based on a dual-core fiber.
\vspace*{2truecm}


\newpage
Localized pulses (solitons) play a central role in a number of nonlinear
physical systems, which are now a subject of a very broad interest [1,2].
Since real systems are always lossy, it is necessary
to have an active element in the system in order to provide
gain compensating the losses. In plasma physics and hydrodynamics,
the gain is provided by intrinsic instabilities of the
system \cite{Lennart}. For solitons in nonlinear optical fibers (NOF's),
an effective way to compensate
for the losses is by use of Erbium-doped amplifiers [2]. Here,
however, one encounters a fundamental problem: if the active element is
uniformly distributed along the length of the system, it automatically renders
the zero solution unstable, thus lending an instability to the whole soliton.
A commonly known model which demonstrates this property is the complex cubic
Ginzburg-Landau (GL) equation, which in many cases, and especially in
application to NOF's, may be regarded as a perturbed nonlinear Schr\"{o}dinger
(NLS) equation \cite{Lennart}:
\begin{equation}%%(1)
iu_t+\frac{1}{2}u_{xx}+|u|^2u\,=\,
i\gamma_0u+i\gamma_1u_{xx}-i\gamma_2|u|^2u\,,
\end{equation}%%(2),
where $u(x,t)$ is an envelope function (e.g., of electromagnetic waves
in the NOF's), $t$ and $x$ are the spatial and temporal variables
(in the NOF's their physical meaning is reversed),
$\gamma_0$ is the gain, while the coefficients $\gamma_1$ and $\gamma_2$
account
for the dispersive and nonlinear losses, and
all the $\gamma$'s are assumed non-negative. A well-known fact is that Eq. (1)
admits an exact solitary-pulse solution, which in the limit of a vanishing
right-hand side of Eq. (1) goes over into the soliton of the
NLS equation \cite{Lennart}. Regarding the pulses in the model (1)
as perturbed NLS solitons, it has been also demonstrated that they are able
to form two-soliton and multisoliton bound states, which are stable against
disturbances of the separation and phase differences between the solitons
\cite{BS}. However, it is obvious that, due to $\gamma_0>0$,
the trivial solution $u=0$ is unstable in this model, hence an isolated
pulse as a whole is unstable too (it was demonstrated numerically that locally
stable pulses are possible in Eq. (1) with {\it negative} $\gamma_0$ and
$\gamma_2$ \cite{Kramer}; we do not consider this case here as the model is
then globally unstable). This circumstance
does not render the pulses meaningless objects - they may be {\it
effectively} stable when the system is short enough, or the evolution is
considered at finite times. In the general case, however, development of the
instability leads to a dynamical chaos \cite{review2}. An example is
the so-called ``dispersive chaos'' experimentally
observed in the binary-fluid convection \cite{Kolodner1}.

A problem of the fundamental interest is to find sufficiently simple physical
models which can produce totally
stable pulses. A known example is the driven damped NLS equation \cite{KN},
which has various applications, including nonlinear fiber optics \cite
{St}. However, in that model the
pulses are not truly localized, being supported by an oscillating background.
Experimentally, absolutely stable localized pulses of traveling-wave
convection were discovered beneath the instability onset in narrow channels
filled with a binary fluid heated from below \cite{convection}.
A distinctive feature of a model supporting stable
pulses is bistability, as, being stable, the localized pulses must coexist
with the stable trivial solution. The simplest possibility to provide for
the bistability is to introduce a {\it quintic} GL equation \cite{quintic}.
In terms of nonlinear fiber optics, this equation models a
nonlinear amplifier \cite{nonlin.ampl}, which can be built, e.g., as a
combination of the linear amplifier and a saturable absorber with an
instantaneous response. In this case, the quintic equation can be
obtained by means of a truncated Taylor expansion of the gain/loss
characterisitic of the medium.

Within the framework of the latter model, the existence of stable
pulses can be demonstrated analytically in two opposite cases: when the
quintic GL equation is
close to the NLS equation \cite{NLS}, and when it is close to the real GL
equation \cite{GL}. For intermediate values of the parameters, existence
of the stable pulses in the model was demonstrated numerically
\cite{FT}. Stable localized pulses for which the quintic equation seems to be
an appropriate phenomenological model were observed in the form of
subcritical pulses in the traveling-wave convection in narrow channels
\cite{convection}. Similar pulses were produced by numerical simulations
of the full system of the corresponding two-dimensional hydrodynamic
equations \cite{simulation}. However, the quintic GL equation proves to
be, even at the phenomenological level, a crude model for the pulses
observed in the experiment. The most essential feature which finds no
explanation in the framework of this model is a very low velocity of the
pulse in the laboratory reference frame \cite{velocity}.
As it was demonstrated by Riecke \cite{Hermann}, an essential improvement
of description of the experimentally observed pulses is achieved within
the framework of a model coupling the {\it cubic} GL equation (actually,
a couple of such equations for counterpropagating waves) to an additional
equation for a real variable, which is a ``mean field'', representing
in this problem a concentration field. It is well known that coupling of
the GL equation to a real mean-field mode can drastically change the
dynamics of the GL model \cite{MF}.

In this work, we aim to put forward a model of another type supporting
absolutely stable solitons.
Actually, the model will generate two solitons, one stable and one unstable,
the unstable one being a separatrix between attraction domains of the stable
soliton and of the stable trivial solution. It will also be demonstrated that
the model may support stable bound states of solitons.

This model describes a dual-core
NOF (also known as a directional coupler \cite{coupler}), in which one core
is active (i.e., doped  with Erbium), while the other is passive
(undoped). The couplers have attracted a lot of attention due to the
possibility
of applications in photonics
\cite{Wabnitz}. In the works \cite{Herb1} and \cite{Herb2}, a
coupler with one active core was proposed. It was
demonstrated that, adding the second passive core with loose ends to the
usual doped fiber used in the soliton-generating lasers, one can essentially
improve quality of the generated solitons: while the soliton, being a
strongly nonlinear object, remains essentially confined to the active core,
linear noise can readily couple to the passive core, where
it is radiated away through the loose ends. This
provides effective filtering of the noise. Independently, it the
work \cite{Aussie} it was proposed to use a coupler for separating
noise from the soliton in an existing pulse. The basic physical idea
proposed in Ref. \cite{Aussie} is the same as in Ref. \cite{Herb1}:
the soliton keeps itself in
the core in which it is propagating, while the noise easily
couples to the second core. It was assumed in Ref.
\cite{Aussie} that the second
core was lossy, so that the noise tunneling to that core would be
killed there by the losses. An important result obtained both
analytically and numerically in Ref. \cite{Aussie} is that the best efficiency
of the filtering would be attained not at very strong losses in the
second core,
but in the case when the loss coefficient is close to the coupling constant
between the two cores.

The numerical simulations reported in Refs. \cite{Herb1} - \cite
{Aussie} demonstrate that the soliton seems very stable in these models
(moreover, in Ref. \cite{Herb2} it was demonstrated that the model produced
a robust soliton even in the case when dispersion in the active core was
{\it normal}, i.e., the solitons could not exist in the usual NLS equation).
These observations suggest to analyze existence and stability of solitary
pulses
in the two-component models. A remarkable fact is that, as it will
be shown below, this problem can be consistently solved in a fully analytical
form.

We adopt the following model, which is a combination of
the more special ones from Refs. \cite{Herb1} and \cite{Aussie} (unlike Eq.
(1), here we will use the standard ``optical'' notation for the spatial and
temporal variables $z$ and $\tau$):
\begin{equation}%%(2)
iu_z+\frac{1}{2}u_{\tau\tau}+|u|^2u-i\gamma_0u-i\gamma_1u_{\tau\tau}+\kappa
v=0,
\end{equation}%%(2)
\begin{equation}%%(3)
iv_z+\frac{1}{2}v_{\tau\tau}+|v|^2v+i\Gamma_0v+\kappa u=0,
\end{equation}%%(3)
where the variables $u$ and $v$ are envelopes of the electromagnetic waves in
the active and passive cores, the coefficients $\gamma_0$ and
$\gamma_1$ are the same as in Eq. (1), the nonlinear dissipation
is neglected (it can be easily restored),
$\kappa$ is the coupling constant, and $\Gamma_0$
is the loss coefficient in the passive core. An additional loss
coefficient $\Gamma_1$ in the passive core, similar to $\gamma_1$ in the active
one, may be added to the model. However, the additional lossy term
would render the following analysis more cumbersome without making any
essential change. Contrary to this, it will be demonstrated
that keeping the term $\sim\gamma_1$ in Eq. (2) is necessary.

As concerns physical applications of this model, it was already mentioned
that it was closely related to the ones describing a soliton-generating laser
\cite
{Herb1} (but in that case, $\Gamma_0=0$) and a time-domain fiber filter
\cite{Aussie}
(however, $\gamma_0=\gamma_1=0$ in the latter model). Very recently, it was
demonstrated
that the model based on Eqs. (2) and (3) in their full form finds another
practically
important application: it is a basis for a new type of a nonlinear optical
amplifier
\cite{MPC}.

First of all, we will consider stability of the trivial solution $u=v=0$.
Inserting into linearized Eqs. (2) and (3) $(u,v)\sim\exp(\sigma z-i\omega t)$,
one can immediately obtain the relation between the instability growth rate
$\sigma$ and the perturbation frequency $\omega$:
\begin{equation}%%(4)
\tilde{\sigma}^2+\left(\Gamma_0-\gamma_0+\gamma_1\omega^2\right)\tilde{\sigma}
+\kappa^2-\Gamma_0\gamma_0+\gamma_1\Gamma_0\omega^2=0,\;
\tilde{\sigma}\equiv \sigma+\frac{1}{2}i\omega^2.
\end{equation}%%(4)
The stability condition
for the trivial solution implies that ${\rm Re}\,\tilde{\sigma}\leq 0$
at all real $\omega$, which is equivalent to demanding that the
free term and the coefficient in front of the linear term in Eq. (4) are always
$\geq 0$. Eventually, this amounts to
\begin{equation}%%(5)
\gamma_0<\Gamma_0<\kappa^2/\gamma_0.
\end{equation}%%(5)
Obviously, a necessary condition following from Eq. (5) is $\gamma_0<|\kappa|$,
i.e., the coupling must be stronger than the gain. Notice that the coefficient
$\gamma_1$ does not appear in Eq. (5).

Now, let us consider evolution of the soliton in the system of Eqs. (2) and
(3).
We will
treat this problem perturbatively, assuming that the coupling, gain, and losses
are all small perturbations to the NLS equation, although different
perturbations may have different orders of smallness. Actually,
we will assume that the gain and losses in the active core are
essentially weaker than the coupling between the cores, while the losses
in the passive one may be comparable to the coupling. Anyway, the
perturbative treatment of all these terms is quite reasonable in
application to the NOF's.

The soliton has two nontrivial parameters, viz., the amplitude and velocity.
The simplest way to derive perturbation-induced
evolution equations for these parameters is to use the so-called balance
equations for the energy and momentum of the
soliton [1]. In our model, the evolution equation for the velocity will
be exactly the same as in the model (1), which describes the
process of ``braking'' of the soliton by the linear frequency-dependent
losses [1], therefore
we will not consider this equation, and will simply set the velocity equal to
zero. Then, in the zeroth-order approximation, the soliton resides only in the
first core and has the form
\begin{equation}%%(6)
u=\eta\,{\rm sech}(\eta\tau)\,e^{i\phi(z)},
\end{equation}%%(6)
where $\frac{d\phi}{dz}=\frac{1}{2}
\eta^2$, and $\eta$ is the soliton's amplitude. In the next
approximation, one seeks for a component (``shadow'') of the soliton in the
second core [1]. Obviously, it has the form $u(z,\tau)=V(\tau)\exp(i\phi(z))$,
where the real function $V(\tau)$ is determined by the equation following
from Eqs. (3) and (6):
\begin{equation}%%(7)
\frac{d^2V}{d\tau^2}-\eta^2V=-2\kappa\eta\,{\rm sech}(\eta\tau)
\end{equation}%%(7)
(at this step, we neglect the lossy term in Eq. (3) which enters only at
the next order). Finally, the slow evolution of the soliton's amplitude under
the action of the perturbations is determined by the above-mentioned balance
equation for the quantity which plays the role of energy in nonlinear optics:
\begin{equation}%%(8)
N\equiv \int_{-\infty}^{+\infty}\left( |u(\tau)|^2+|v(\tau)|^2\right)d\tau.
\end{equation}%%(8)
An exact balance equation following from Eqs. (2) and (3) is
\begin{equation}%%(9)
\frac{dN}{dz}=2\gamma_0\int_{-\infty}^{+\infty}|u(\tau)|^2d\tau-2\gamma_1
\int_{-\infty}^{+\infty}|u_{\tau}(\tau)|^2d\tau-2\Gamma_0\int_{-\infty}
^{+\infty}|v(\tau)|^2d\tau.
\end{equation}%%(9)

Now, one should find $V(\tau)$ from Eq. (7) and insert it into the last term
on the right-hand side of Eq. (9), while a contribution from $v$ to the
left-hand side (see Eq. (8)) may be neglected in the lowest nontrivial
approximation. Eq. (7) can be solved by means of the Fourier
transformation, which leads to the following integral representation for
$V(\tau)$:
\begin{equation}%%(10)
|V(\tau)|=|\kappa|\int_{-\infty}^{+\infty}{\rm sech}\left(\eta\tau^{\prime}
\right)e^{-\eta|\tau-\tau^{\prime}|}d\tau^{\prime}.
\end{equation}%%(10)
Inserting this expression into the last term of Eq. (9), one can explicitly
calculate all the integrals, which eventually yields the evolution equation
sought for:
\begin{equation}%%(11)
\frac{d\eta}{dz}=2\gamma_0\eta-\frac{2}{3}\gamma_1\eta^3-C\kappa^2\Gamma_0
\eta^{-3},
\end{equation}%%(11)
where $C\equiv \frac{1}{6}\pi^2+\zeta(3)\approx 2.845$.
The formal singularity of the last term in Eq. (11)
at $\eta\rightarrow 0$ is fictitious,
as this expression is irrelevant at very small $\eta$, see below.

It is straightforward to see that Eq. (11) gives rise to two physical $(\eta^2
>0)$ fixed points, provided that
\begin{equation}%%(12)
\gamma_0^3\,>\,(3C/8)\kappa^2\Gamma_0\gamma_1^2,
\end{equation}%%(12),
and to no fixed points in the opposite case. Thus, (12) is the necessary and
sufficient condition for existence of the solitons in the considered
model. It is, of course, important to check if this condition is compatible
with the other fundamental condition, (5), which is necessary for stability
of solitons in the model. Because
Eq. (5) does not involve the parameter $\gamma_1$, one can
secure the compatibility simply choosing $\gamma_1$ to be small enough.

Next, it is easy to check that, once the condition (12) is met,
the solution with larger $\eta^2$ is stable, and the one with smaller
$\eta^2$ is unstable. It is very plausible that the soliton corresponding to
the
larger root is a completely stable solution in the full model, while the
smaller
root corresponds to an unstable soliton which plays the role of a separatrix
between the stable soliton and the stable trivial solution.

Usually, existence of solitons is related to modulational instability of
continuous wave (cw) solutions [2]. The cw modulational (in)stability in the
present model will be considered in detail elsewhere. However, in the regime
which
is akin to the case considered in this work, i.e., when the coefficients of the
gain and loss are small, the coupling constant is small too, and the field
in one core ($v$) is therefore much weaker than in the other ($u$), the
modulational
instability in the present model is, evidently, close to that in the single
NLS equation. Thus, as well as in the usual NLS equation, one may qualitatively
regard the solitons in the present model as pulses produced by the
modulational
instability of the cw.

It is now relevant to discuss conditions guaranteeing application of the
perturbation theory to this problem. The ``primary'' conditions are $\gamma_1
\ll 1,\,|\kappa|\ll \eta^2$, and $\Gamma_0\ll \eta^2$. Since, in a
typical case, $\eta^2\sim\gamma_0/\gamma_1$, the final set
of the applicability conditions takes the form
\begin{equation}%%(13)
\gamma_1\ll 1,\,\gamma_1\Gamma_0\ll \gamma_0,\, \gamma_1|\kappa|\ll\gamma_0.
\end{equation}%%(13)
Obviously, these conditions are compatible with the underlying inequalities
(5) and (12).

When formulating the model, we have omitted the dispersive loss
term $i\Gamma_1u_{\tau\tau}$
in Eq. (3). If kept, it will generate an additional term $\sim -\kappa^2
\Gamma_1\eta^{-1}$ in Eq. (11). It is easy to check that the latter
term will not produce any qualitative difference in properties of the fixed
points. On the other hand, if the term $\sim \gamma_1$
is omitted in Eq. (2),
the result will be disastrous: there will remain a single unstable
fixed point. Actually, the model is globally
unstable in that case. At last, if the nonlinear lossy term $\sim\gamma_2$
(the two-photon absorption, in terms of the nonlinear optics) is
added to Eq. (2) (cf. Eq. (1)), it will merely renormalize
the coefficient $\gamma_1$ in the final results displayed above.

Now, we will briefly consider bound states (BS's) of the solitons in
this model, following the lines of Ref. \cite{BS}. The BS's may
exist due to the fact that the small linear terms accounting for the gain and
dissipation render
solitons' tails oscillatory, which, in turn, gives rise to local minima in
the effective potential of the soliton-soliton interaction, produced by
overlapping of the ``head'' of each soliton with the tail of the other
one. However, in
the framework of the perturbation theory, the BS's are fragile, although
stable:
the distance between the solitons in the BS is large, and, accordingly,
the corresponding binding energy is exponentially small. Nevertheless,
existence and stability of the BS's predicted by the perturbation theory
was confirmed by direct numerical simulations \cite{Cai} of the driven
damped NLS equation \cite{KN} (see also \cite{St}); recently, this
prediction was
also confirmed, with a fairly good accuracy, for the cubic GL equation (1)
\cite{Jena}. In the present model, the BS's can be rendered
more robust by increasing the dissipative constant $\Gamma_0$ in the passive
core. Therefore, we will consider the case
$\Gamma_0\gg\gamma_0,\gamma_1\eta^2$, which is
compatible with all the above conditions necessary for existence of the stable
solitons. In this case, we again consider
linearized Eqs. (2) and (3); however, instead of the plane-wave solution
leading to the dispersion equation (4), we are now interested in an
exponentially decaying
solution describing the soliton's tail:
\begin{equation}%%(14)
(u,v)\,\sim \,\exp(-\eta|\tau|+ i\chi|\tau|+iqz)\,,
\end{equation}%%(14)
where we consider $\eta$ as a given soliton's inverse size
(cf. Eq. (6)), while $\chi$ and $q$ must be found. The linearized
equations immediately yield (with regard to the assumed dominance
of $\Gamma_0$):
\begin{equation}%%(15)
\chi=\Gamma_0/2\eta,\,\left(q-\frac{1}{2}\eta^2+\frac{1}{2}\chi^2\right)^2=
\kappa^2-\frac{1}{4}\Gamma_0^2.
\end{equation}%%(15)
According to Ref. \cite{BS}, the minimum separation $T$ between
solitons in the BS is determined by the coefficent $\chi$ in Eq. (14):
$T=\pi/2\chi=\pi\eta/\Gamma_0$, where we have made use
of Eq. (14). On the other hand, the second
relation in Eq. (15) imposes a fundamental limitation
$\Gamma_0<2|\kappa|$, otherwise the soliton simply does not exist. This,
in turn, leads to a limitation on the minimum separation between the bound
solitons: $T>\pi\eta/2\kappa$.

In conclusion, we have demonstrated that a simple analytically tractable
model, based on the linearly coupled cubic GL equations, admits
a fully stable soliton coexisting with the stable trivial state. The
model finds a direct physical realization in terms of the fiber
laser and suggests a way to stabilize soliton generation in the laser.

This research was partially supported by the National
Science Foundation through the Center for Ultrafast Optical Science at the
University of Michigan, Ann Arbor.

\end{document}